# Rational layers in the low temperature expansion of the three-dimensional Ising model

Ran Huang

Centre of Biomedical Engineering and Technology, Yiwu Research Institute of Fudan University, Yiwu 322000, China

* Correspondence: huangran@fudan.edu.cn

## Abstract

A low-temperature expansion orders Ising droplets by wall area, but droplets with different transverse spans can occur at the same order. We instead organize the zero-field simple-cubic model by the excess above the minimum wall area compatible with two prescribed spans. We prove that every fixed excess layer, including all logarithmic cover corrections and Euler inversion, has a rational bivariate generating function with denominator factors restricted to $1 - x$, $1 - y$ and $1 - x - y$. The proof compresses repeated coordinate cuts into bounded decorations and assigns each unbounded interval to one monotone chain. A finite controller then enforces induced contacts and arbitrary cover-color reuse. We explicitly assemble the complete layers of excess zero through three over all widths, attributing their known geometric constituents to earlier polyomino enumeration. Exact lattice checks corroborate the formulas. A certified low-temperature comparison shows a modest hybrid accuracy gain at one sampling point and higher evaluation cost. The result establishes an exact layer structure; summation over unbounded excess remains a separate problem.

## Significance Statement

The three-dimensional Ising model is a cornerstone of phase-transition theory, but its exact solution remains unknown. We organize its low-temperature expansion into layers defined by boundary area above a minimum set by spatial extent. Each layer contains structures of arbitrarily large extent in two directions, yet we prove that its generating function is a ratio of polynomials with only three possible denominator factors. We evaluate the first four layers explicitly, including the corrections required to obtain their thermodynamic contributions. This organization reveals exact mathematical structure hidden within conventional expansions and establishes a systematic foundation for calculating further layers and investigating how increasingly complex spin clusters contribute to the collective behavior of a three-dimensional system.

## 1 Introduction

A large droplet and a thin, extended droplet need not have comparable effects on a low-temperature expansion. Their wall areas set the same energetic penalty when the areas coincide, yet their spans and internal geometry can be very different. A wall-area coefficient combines these geometries before their separate spatial organization becomes visible. The present work retains two transverse spans and asks a different question: what remains complicated when the wall area exceeds its minimum at those spans by only a fixed amount?

The exact two-dimensional zero-field solution and its spinor reformulation provide a complete evaluation of the thermodynamic free energy [1, 2]. For the cubic model, finite-lattice expansions instead recover exact coefficients by canceling boundary and span contributions [3, 4], while recursive state enumeration has generated long low-temperature series [5]. These methods establish a strong computational baseline. Our purpose is to resolve a structural class containing unbounded transverse widths rather than to extend the record length of an area series.

The relevant geometric foundation is the enumeration of animals inscribed in a rectangle or prism. Minimum-area planar animals and the next area class have explicit formulas [6]. Subsequent work treats index-two planar animals [7], and minimum-volume three-dimensional animals have rational generating functions, including an exact height-two lifting identity [8]. Pellerin gives a systematic account of the planar index and its relation to holes and faces [9]. The link between recursive geometry and rational generating functions is a general principle in exact animal enumeration [10]. These precedents account for several of the low-excess geometric ingredients used below.

The additional issue for an Ising pressure is that a support count is not a pressure coefficient. The logarithm introduces ordered covers, overlapping supports, returning colors and signed corrections. Euler inversion then redistributes repeated areas. We prove rationality for the complete fixed-total-excess layer, retaining these operations. The central step is a canonical embedding construction in which every unbounded coordinate interval has one owner. This removes the length-matching constraints that would otherwise require a coefficient diagonal or an unevaluated growing transfer matrix.

The result has two levels. For each fixed total excess, it gives a finite construction in a specified rational function ring. For the first four layers, explicit formulas yield complete all-width evaluations. A worked third layer displays how known planar counts, nonplanar lifts and cover corrections combine. We then compare the resulting partial resummation with ordinary area truncation. The comparison distinguishes exactness of the retained layers from accuracy and cost of the resulting pressure approximation.

# 2 Formal pressure and the layer theorem

## 2.1 Spans wall area and connected supports

Let the lattice be $\mathbb{Z}^3$, with uniform ferromagnetic coupling $K > 0$ and zero field. For a finite vertex set $X$, write $e(X)$ for its number of induced nearest-neighbor edges and $P(X) = 6|X| - 2e(X)$ for its wall area. With a plus exterior and the aligned-bond energy removed, the finite-set partition polynomial is

$$\Xi_T(q) = \sum_{X \subseteq T} q^{P(X)}, q = e^{-2K}. \quad (1)$$

The physical pressure per site is $p_3(K)$ and its vacuum correction is $g(q) = p_3(K) - 3K$. On arbitrary finite sets we use the vertex Möbius weight

$$\ell(S) = \sum_{T \subseteq S} (-1)^{|S|-|T|} \log \Xi_T(q). \quad (2)$$

The identity $\log \Xi_\Lambda = \sum_{S \subseteq \Lambda} \ell(S)$ is formal and exact. If $S$ is disconnected, the partition polynomial factors across its components and $\ell(S) = 0$. For connected $S$, let $a, b, c$ be its coordinate widths. Translation classes are counted once; rotations and reflections remain distinct. Summing $\ell(S)$ over classes with exact transverse widths $a, b$ defines $J_{ab}(q)$. Coefficientwise finite-volume density limits give $g(q) = \sum_{a,b} J_{ab}(q)$ as a formal low-temperature series.

Expand the span pressure into Euler factors [11], and define the reference area and total excess layer by

$$J_{ab}(q) = -\sum_{m \geq 1} E_m(a,b) \log(1 - q^m), m_0(a,b) = 4a + 4b - 2. \quad (3)$$

$$R_J(x,y) = \sum_{a,b \geq 1} E_{m_0(a,b)+2J}(a,b) x^{a-1} y^{b-1}, Q_J(t) = R_J(t,t). \quad (4)$$

Here $J$ is a nonnegative integer layer index, whereas a support has geometric excess $j = (P(S) - m_0)/2$. Cover corrections can make $J > j$. The substitution $x = y = t$ is ordinary substitution; it does not extract the coefficients of $x^n y^n$.

## 2.2 Rationality at every fixed excess

**Theorem 1. For every fixed integer $J \geq 0$, both the pressure layer before Euler inversion and the Euler layer $R_J$ belong to**

$$\mathcal{R} = \mathbb{Q}[x, y, (1-x)^{-1}, (1-y)^{-1}, (1-x-y)^{-1}]. \quad (5)$$

The proof gives a finite construction depending on $J$. Consequently, the only possible finite poles of $Q_J(t)$ are at $t = 1$ and $t = 1/2$. Its sufficiently large width coefficients have the form

$$[t^h]Q_J(t) = A_J(h) + 2^h B_J(h), \tag{6}$$

where $A_J$ and $B_J$ are polynomials with rational coefficients. The theorem fixes the denominator type, not optimal pole orders or the dependence of the numerator on unbounded $J$. A conservative explicit state bound is given in Supplementary Section S2. The construction underlying the theorem is finite, while the supplied executable code implements local controllers and the explicit low layers rather than exhaustive type generation for arbitrary $J$.

# 3 Proof by finite geometry and monotone chains

## 3.1 The bounded geometric part

For a connected support write $n = |S|$, $r = e(S) - n + 1$ and $s = n - (a + b - 1)$. Let $N_{xy}, N_{xz}, N_{yz}$ be projection sizes, and let $I_x, I_y, I_z$ count occupied interval components on all coordinate fibers. Counting exposed bonds on each fiber gives $P/2 = I_x + I_y + I_z$. A connected planar projection with widths $u, v$ has at least $u + v - 1$ vertices. Thus the geometric excess is a sum of six nonnegative integers:

$$j = (I_z - N_{xy}) + (I_y - N_{xz}) + (I_x - N_{yz}) \tag{7}$$

$$+ (N_{xy} - a - b + 1) + (N_{xz} - a) + (N_{yz} - b). \tag{8}$$

In particular $j \geq 0$ and $P \geq 4(a + b + c) - 6$, so $c \leq \lfloor j/2 \rfloor + 1$. To bound the volume beyond minimum transverse span, put $u = N_{xy} - (a + b - 1)$, $v = N_{xz} - a$ and $w = N_{yz} - b$. The distinct height sets at a given $x$ and a given $y$ bound the extra occupancy of a vertical fiber, giving $n - N_{xy} \leq vw$. Since $u + v + w \leq j$,

$$0 \leq s \leq S_j := \max\{j, \lfloor j^2/4 \rfloor\}, r = 2s - j. \tag{9}$$

Call a global transverse cut repeated if at least two support edges cross it. Let $B$ contain all edges on repeated transverse cuts and every vertical edge. Each edge outside $B$ is a bridge, because removing the sole edge crossing its global cut disconnects the support. If $E_x, E_y, E_z$ count directional edges, then

$$(E_x - a + 1) + (E_y - b + 1) + E_z = 3s - j, |B| \leq 6S_j - 2j. \tag{10}$$

Mark every endpoint of $B$ and every vertex of degree other than two. The singleton is marked separately. A minimal subtree of a spanning tree joining four transverse extreme vertices has at most four leaves and at least $a + b - 1$ vertices. At most $s$ vertices lie outside it, so the original graph has at most $4 + s$ leaves. The degree identity then bounds the marked set $M$ and the number $C$ of maximal good chains:

$$|M| \leq 6 + 18S_j - 6j, C \leq 3|M|. \tag{11}$$

Every good chain lies at constant height and is monotone in each transverse coordinate: reversal would cross the same globally unique cut twice. There are no good-chain loops or parallel good chains. Marked vertices are ordered lexicographically by their actual coordinates, and each chain is oriented by its ordered endpoint labels. This makes the decomposition canonical.

## 3.2 Compression and global coordinate ownership

Deleting the unit interval associated with every repeated transverse cut and forgetting height collapses every connected component of $B$ to a point. The abstract contracted graph $T$ is a tree. If $h_x, h_y$ are the numbers of repeated cuts, its number of edges is $(a - 1 - h_x) + (b - 1 - h_y)$. Its connected geometric image has widths $a - h_x, b - h_y$ and therefore at least as many vertices as $T$. Since contraction cannot create additional image points, equality holds and the image is injective. Every cut is used once, so a tree path between image neighbors has length one; the image is also induced. Fig. 1 illustrates this reduction.

Compression alone does not enumerate inverse embeddings. Distinct decorations may share the same coordinate band, and endpoint offsets can couple attachments. We retain these identities in original coordinates. For every marked coordinate, form the events at offsets $-1, 0, +1$. There are at most $3|M|$ event levels per axis. A finite type specifies their weak order and whether each intervening gap has length one or length at least two. The offset events are constrained to lie one unit from their central event. Heights are bounded by the previous lemma.

Within the marked span, a gap crossed by $B$ is a unit gap with at least two such edges and no good-chain owner. Every other gap has exactly one good-chain owner in that axis. In particular, each unbounded gap has a unique owner. A chain with constant coordinate may lie inside another chain's positive projection interval, but that coordinate is an event level and does not introduce a second free length. Different owners therefore never have to match two independently summed unbounded lengths.

### 3.3 Finite probes enforce induced contacts

Any intersection or nearest-neighbor contact between different chains, or between a chain and a marked vertex, occurs on the finite grid of transverse event levels and bounded heights. To see this one axis at a time, two positive chain projection intervals use disjoint cut intervals; they can approach only at endpoints or one unit from endpoints. If one interval is constant, its value is a marked coordinate and the other point must be within one unit. Applying both axis statements gives the probe claim.

A monotone chain has no internal unintended nearest-neighbor edge, since its path distance equals the Manhattan distance. A finite mask recording visited probes is consequently enough to reject interior collisions, visits to unintended marked vertices and additional cross-chain contacts. Probe adjacency depends only on unit-gap flags and height differences. It does not depend on the integer length of a long gap, including the case in which that length is exactly two.

### 3.4 Logarithmic covers have a finite cost budget

Expanding the logarithm before the vertex Möbius sum gives the exact ordered-cover representation

$$\ell(S) = \sum_{k\geq 1} \frac{(-1)^{k+1}}{k} \sum_{X_1\cup\cdots\cup X_k = S} q^{\sum_i P(X_i)}. \tag{12}$$

Every cover member is nonempty; members may be disconnected or repeated. No factor $1/k!$ is introduced. Assign each vertex its nonempty set $A_v \subseteq \{1, \dots, k\}$ of cover labels. Direct bond counting gives the nonnegative excess cost

$$D = \sum_{uv\in E(S)} |A_u \triangle A_v| + \sum_{v\in S} (6 - \deg v)(|A_v| - 1) = 2(J - j). \tag{13}$$

Split every cover member into connected components. Their intersection-or-contact graph is connected because their union is $S$. On merging a new component with the accumulated union, the boundary deficit is at least two, using

$$P(A) + P(B) - P(A \cup B) = P(A \cap B) + 2e(A \setminus B, B \setminus A). \tag{14}$$

The sum of component perimeters equals the sum of cover-member perimeters. There are at least $k$ components, so $D \geq 2(k-1)$ and $k \leq J - j + 1$. Thus both the number of labels and the cost are bounded at fixed $J$. This argument does not assume a connected color class or forbid the reuse of a color after it disappears.

### 3.5 Rational evaluation and reconstruction

For each chain, record its coordinate phase in the common event table, its probe mask, current color, seen colors and cost. A long gap is traversed by entering its interior, making any number of interior steps and then exiting. An $x$ or $y$ step has the corresponding width weight. At an internal chain vertex, an edge from color $A$ to color $A'$ adds $|A \triangle A'| + 4(|A'| - 1)$ to the cost; endpoint vertex terms are charged once in the finite marked core.

Every transition other than a self-loop increases cost or a coordinate phase. A zero-cost repeated state has one unchanged singleton color. Its self-loop weight is therefore only $0, x, y$ or $x + y$. Reverse substitution evaluates each accepted terminal category by

$$F_\sigma = \frac{r_\sigma + \sum_{\sigma \to \tau \neq \sigma} w_{\sigma\tau} F_\tau}{1 - w_\sigma}. \quad (15)$$

All accepted-chain functions belong to $\mathcal{R}$. Their finite products and the finite global probe, degree, color and cost filters remain in the same ring. A real support and cover give a unique type and collection of annotated chain words. Conversely, every accepted word set determines each long-gap length through its owner, reconstructs the common coordinate table, and recovers the induced support and cover. Re-encoding returns the same data. The geometry label is independent of the long lengths: with $L$ total good steps, $n = |M| + L - C$, $e = |B| + L$ and $(a - 1) + (b - 1) = L + h_x + h_y$. The specified geometric excess is therefore preserved. These observations prove the required bijection and rationality of the pressure layer. The complete state specification is in Supplementary Section S2.

Finally, Euler inversion is

$$E_m(a, b) = \sum_{d|m} \frac{\mu(d)}{d} [q^{m/d}] J_{ab}(q). \quad (16)$$

A nontrivial divisor can contribute at $m = m_0 + 2J$ only if $m \geq 2m_0$, which implies $a + b \leq \lfloor (J + 1)/2 \rfloor$. Thus Euler inversion changes only finitely many widths at fixed $J$. It adds a polynomial and preserves $\mathcal{R}$, proving Theorem 1. Partial fractions after $x = y = t$ yield the coefficient form stated above.

# 4 Four explicit layers and the complete third layer

The first layer counts minimum planar animals in the present normalization and reproduces the established formula [6]. The planar one-cycle and one-gap-tree constituents also reproduce known index-one counts [6, 7]. The height-two minimum-volume lift is the weighted planar-tree identity of Goupil and Cloutier [8]. We use these facts as inputs and normalization checks. The complete layer additionally includes logarithmic covers and, when permitted, Euler divisor corrections.

$$Q_0(t) = \frac{1 - 4t + 8t^2 - 6t^3 + 4t^4}{(1 - t)^4 (1 - 2t)}. \quad (17)$$

The complete layers through $J = 3$ have reduced denominators $(1 - t)^{J+4}(1 - 2t)^{J+1}$. This observation is asserted here only for the four explicit functions. Theorem 1 does not require or establish this sharper pattern for arbitrary $J$. Full numerators, geometric derivations and cover accounting are given in Supplementary Sections S3–S5. The low-width coefficients are listed in Table 1.

Table 1. **Complete Euler-layer coefficients at small width degree h = a + b − 2.**

| J | h = 0 | h = 1 | h = 2 | h = 3 | h = 4 |
|---|---|---|---|---|---|
| 0 | 1 | 2 | 6 | 18 | 51 |
| 1 | 0 | −2 | −11 | −46 | −164 |
| 2 | 1 | 8 | 54 | 276 | 1152 |
| 3 | −2 | −18 | −142 | −834 | −3954 |

For geometric excess three, the height bound allows at most two layers. In a plane, the exact relation $j = r + 2g$ separates convex tricyclic supports from one-gap unicyclic supports, where $g$ counts additional row and column interval components. The latter class and the convex bicyclic ingredients have direct antecedents in index-two enumeration [7, 9]. Two-layer excess-three supports have either a duplicated vertex of a convex unicyclic projection or a duplicated adjacent pair in a minimum planar tree. The complete geometric sum is denoted $Z_3$.

Let $D = td/dt$, $N = Q_0$, $U$ be the complete planar one-cycle count, $W_N$ the degree-weighted minimum-tree count and $B_2$ the complete geometric-excess-two bridge count. Including lower-support covers and the primitive correction gives

$$Q_3 = Z_3 - B_2 + \frac{1}{2}(D+1)(D-6)U \quad (18)$$

$$-\frac{1}{6}D(D-1)(D-2)N - \frac{1}{2}W_N - \frac{1}{2}. \quad (19)$$

The four nongeometric terms are, respectively, one bridge split, the second cover coefficient of a single-square support, disjoint tree covers plus the first overlap, and the only nontrivial Euler divisor at this order. Exact rational assembly yields

$$Q_3(t) = \frac{P_3(t)}{(1-t)^7(1-2t)^4}. \quad (20)$$

The degree-15 numerator is supplied in Supplementary Section S5. Its nonzero leading residues are 60 at $t = 1$ and $-27/8$ at $t = 1/2$. Hence both displayed pole orders are exact. Fig. 2 separates the geometric classes, cover contributions and the polynomial-plus-exponential width behavior. These statements follow from finite rational identities, without coefficient fitting.

Exhaustive enumeration of 50,148 fixed planar animals through ten vertices and direct projection lifts reproduce the first seven complete $Q_3$ coefficients. A separate plus-boundary spin-transfer calculation, followed by span differences and Euler inversion, reproduces the first four width coefficients of all four layers through wall area 24. These checks test implementations and low-order consequences of the analytic classification; the all-width statements rest on the classification and assembly proofs.

# 5 Certified pressure evaluation and its cost

Put $B_E(q) = q^6 \sum_{J\geq 0} q^{2J} Q_J(q^4)$. Repeating Euler primitives reconstructs the pressure as $g(q) = \sum_{\ell\geq 1} B_E(q^\ell)/\ell$. Supplementary Section S6 derives a contour majorant using the polymer expansion criterion [12]. At $r = 43/160$, its absolute pressure budget is $B = 1/200$; span projection and Euler inversion give $C = B/(1-r^6)$ for the joint absolute Euler norm. Thus all widths, layers and repeats may be regrouped absolutely for $0 \leq q \leq 1/4$.

$$|g(q) - g_N(q)| \leq \frac{C}{1-q^{2N+8}} \left(\frac{q}{r}\right)^{2N+8}. \quad (21)$$

This is a sufficient bound on the omitted layers, not a value of their actual contribution. It does not imply that a high $N$ has been computed. Repeated powers are evaluated through $\ell = 6$, with the remaining repeat tail bounded separately.

We compare four approximations at $q = 1/10, 1/6, 1/4$: the complete pressure polynomial $P_{26}$ through area 26; the primitive-area approximation $A_{26}$ with its repeats; the all-width layer approximation $G_3$ through excess three; and a hybrid $H_{3,26}$ that adds every primitive of area at most 26 omitted by $G_3$. Their retained sets differ, as shown in Fig. 3a. In particular, $G_3$ is not a higher-order area series: it already omits one primitive at area 14, represented by a vertical three-site rod. Accordingly,

$$g(q) - G_3(q) = q^{14} - 4q^{16} + 32q^{18} + O(q^{20}). \quad (22)$$

The common norm supplies rigorous certificates for all four approximations (Table 2). A second, tighter reference uses the published per-site partition series through $u^{26}$ [4], with $u = q^2$. We take its exact formal logarithm to obtain a pressure polynomial $P_{52}$, rather than using partition coefficients as pressure coefficients. A norm tail surrounds that center by an interval. Statements based on this interval are conditional on the published higher coefficients and the contour norm; $P_{52}$ itself is not the exact pressure.

Table 2. **Absolute error certificates from the common norm, including the repeat tail. Displayed upper bounds are rounded upward.**

| q | $P_{26}$ | $A_{26}$ | $G_3$ | $H_{3,26}$ |
|---|---|---|---|---|

| q | $P_{26}$ | $A_{26}$ | $G_3$ | $H_{3,26}$ |
|---|---|---|---|---|
| 1/10 | $4.392 \times 10^{-15}$ | $4.393 \times 10^{-15}$ | $4.505 \times 10^{-9}$ | $4.393 \times 10^{-15}$ |
| 1/6 | $7.151 \times 10^{-9}$ | $7.154 \times 10^{-9}$ | $5.748 \times 10^{-6}$ | $7.154 \times 10^{-9}$ |
| 1/4 | $6.094 \times 10^{-4}$ | $6.097 \times 10^{-4}$ | $1.678 \times 10^{-3}$ | $6.097 \times 10^{-4}$ |

At $q = 0.1$, the conditional reference places the absolute errors of $P_{26}$ and $H_{3,26}$ near $2.2 \times 10^{-24}$ and $1.7 \times 10^{-24}$, respectively. Conservative interval comparison gives a hybrid reduction between 17.7% and 22.7%. The error of $G_3$ is about $9.63 \times 10^{-15}$ and is demonstrably larger. At $q = 1/6$ the reference also separates $G_3$ from the much more accurate complete area polynomial, but does not rank the hybrid against $P_{26}$. At $q = 1/4$ the reference is too broad to resolve their small center differences. Fig. 3b,c keeps certificates, conditional intervals and unresolved comparisons distinct.

Every online evaluator uses exact rational arithmetic in an isolated process, with five warmed batches and the same repeat cutoff. The hybrid costs substantially more per evaluation than the sparse area polynomial; no computational advantage is demonstrated (Fig. 3d). Separate preparation replays start from different inputs—proved rational components for $Q_3$, actual finite-support enumeration for the area coefficients—and therefore do not define an end-to-end speed comparison. Source data include both timings and process memory, with interpreter overhead retained.

## 6 Discussion

The rational layer structure separates spatial extension from geometric excess. At fixed excess, repeated cuts, height excursions and branching complexity are bounded. Unbounded extension survives only along monotone chains whose admissible repetitions produce three denominator factors. Logarithmic covers preserve this structure because every color change or nonsingleton occupation consumes a nonnegative budget. Euler inversion preserves it because nontrivial divisors are confined to finitely many widths.

This separation also identifies the remaining difficulty. A compact cube of side $L$ has geometric excess $3L^2 - 4L + 1$; the smallest $2 \times 2 \times 2$ cube occurs at excess five. Thus arbitrary three-dimensional droplets require unbounded excess even though every fixed layer contains unbounded widths. The present construction does not replace that hierarchy by one fixed-size equation. Similarly, the fixed-layer poles are width-counting singularities, not the physical critical point. With $t = q^4$, the pole $t = 1/2$ corresponds to $q = 2^{-1/4}$, whereas a numerical critical benchmark is $q_c \simeq 0.64191$ [13]. The physical singularity requires information from the sum over layers.

The finite-state construction is deliberately conservative. Its purpose is to establish a rational function type without hiding length matching, contact exclusions or color reuse. An efficient enumeration of the finite types may sharpen the state bounds and automate higher-layer evaluation. The current benchmark makes clear that these algorithmic questions cannot be settled from all-width exactness alone.

## 7 Conclusion

Organizing the cubic Ising low-temperature expansion by excess above the minimum area at fixed transverse spans yields rational complete layers. A canonical finite embedding and cover controller proves that only three bivariate denominator factors occur. Explicit evaluation through excess three illustrates the full passage from geometry to pressure and Euler primitives. Certified comparisons show both the information retained by this resummation and its present computational cost. The resulting exact partial structure provides a defined basis for studying the remaining sum over unbounded excess.

## Data and code availability

The accompanying reproducibility package contains exact-arithmetic checks, lattice enumerators, the local chain controller, figure-generation code and the measured benchmark data. A single runner reproduces the checks and both preparation replays. Supplementary Section S7 specifies dependencies, commands and the extent of each computation. The higher-order reference coefficients are transcribed from table 2 of Ref. [4].

## Artificial Intelligence (AI) Disclosure

OpenAI Codex assisted with code plot and manuscript polish. The authors are responsible for the scientific content and its final review.

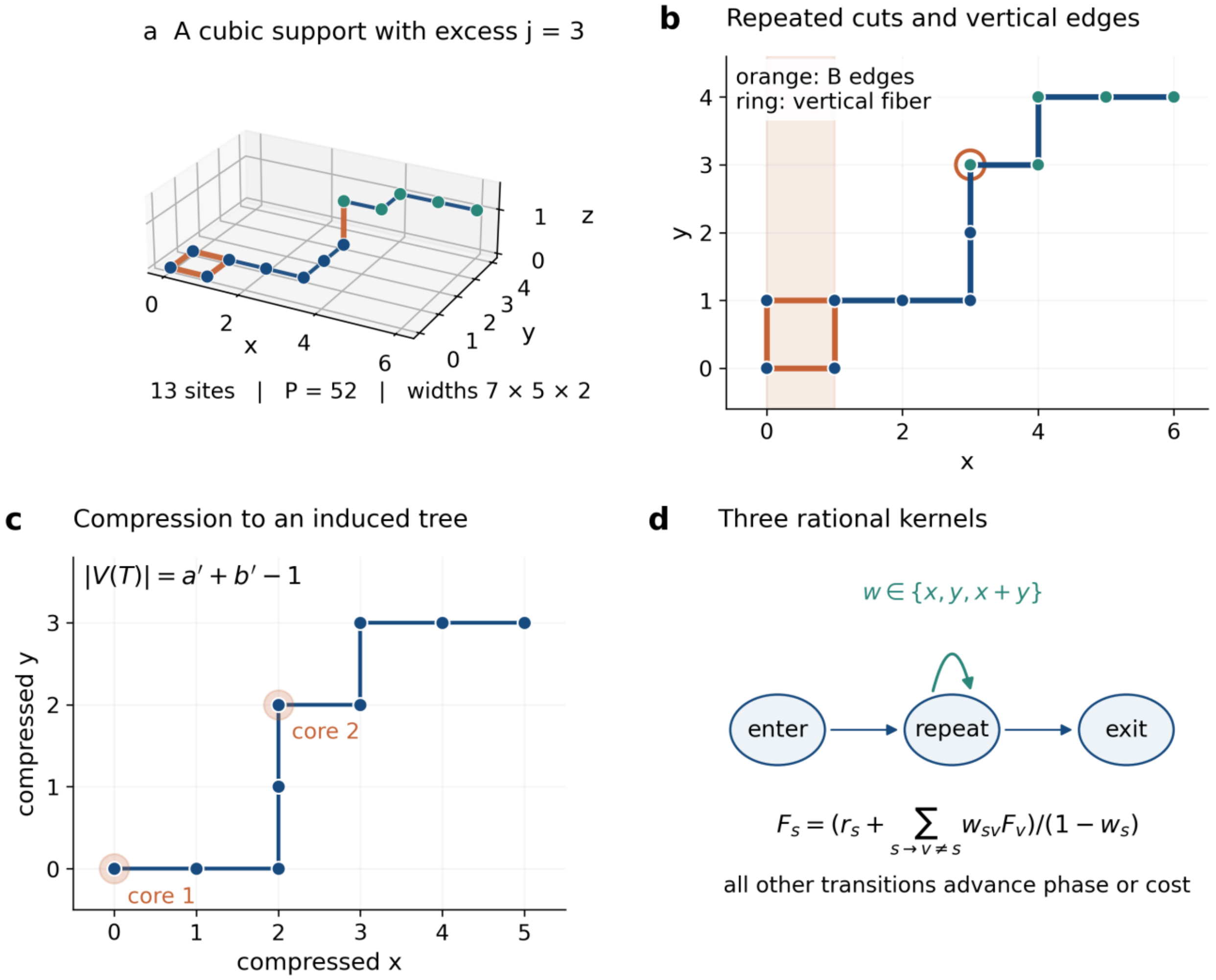

Figure 1. Bounded complexity and unbounded extension. (a) An actual induced cubic support, with orange edges on repeated transverse cuts or in the vertical direction. (b) Its transverse projection retains the identities of repeated cuts and vertical fibers. (c) Compression produces an injectively embedded induced minimum planar tree; shaded vertices indicate the contracted decorations. (d) Schematic local chain evaluation. Repeating a zero-cost singleton color generates only x, y or x + y self-loop weights; all other transitions advance spatial phase or cost. The full controller also retains probe masks, colors and endpoint constraints. The geometry in panels a–c is generated and checked from literal lattice coordinates.

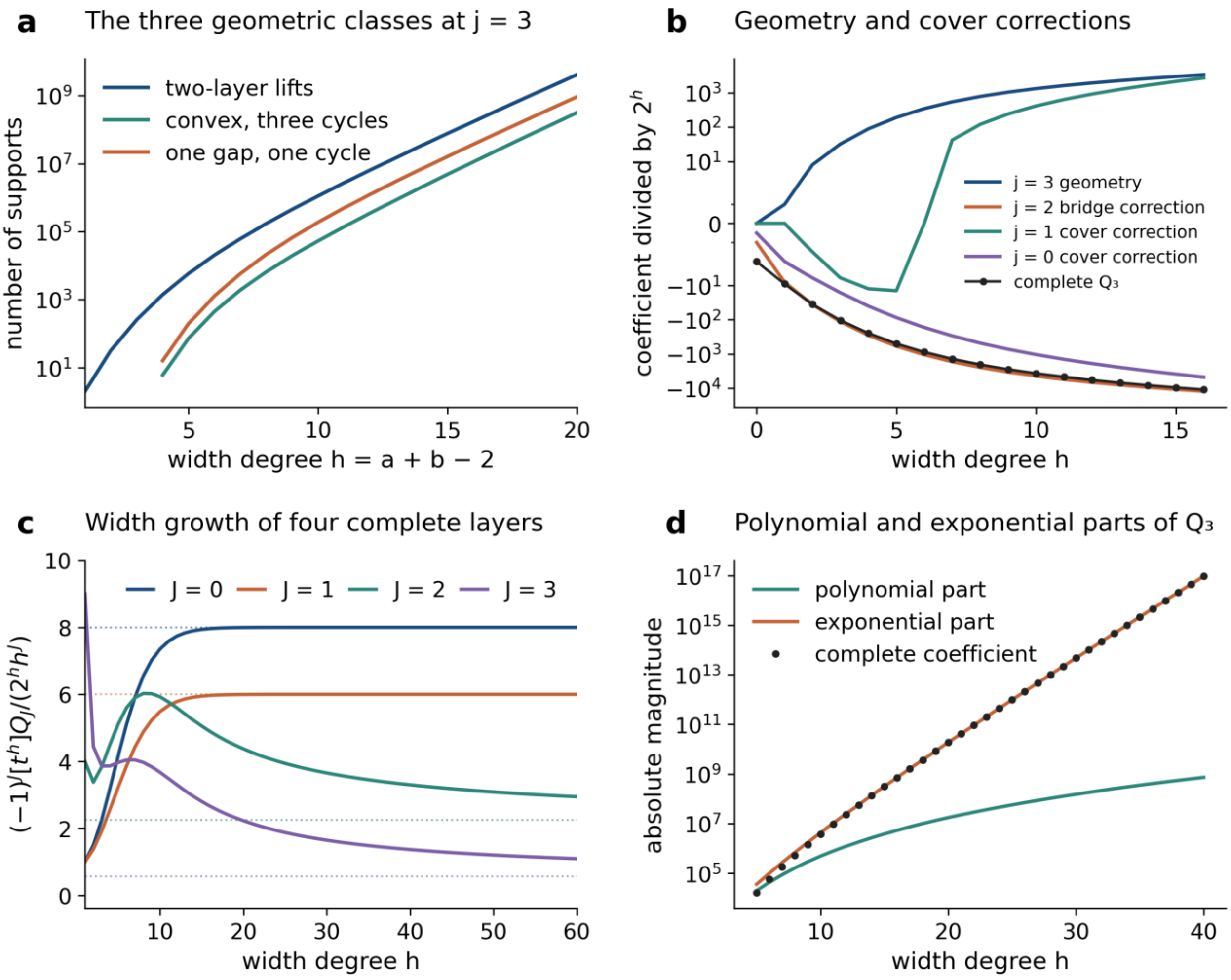

Figure 2. Exact assembly and width behavior of the complete third layer. (a) Counts of the three exhaustive geometric-excess-three classes. (b) Signed contributions of geometric excess three, bridge cuts, one-cycle covers and minimum-tree covers, scaled by $2^h$; the complete Euler layer includes the additional constant primitive correction. (c) Width coefficients for the four explicit layers, normalized by their exact exponential-polynomial leading behavior. Dotted levels are computed from these four rational functions only and make no all-J pole claim. (d) The exact $Q_3$ coefficient for $h \geq 5$ splits into a polynomial and $2^h$ times a polynomial. Magnitudes are shown; the exponential part is negative. All plotted coefficients come from exact rational identities.

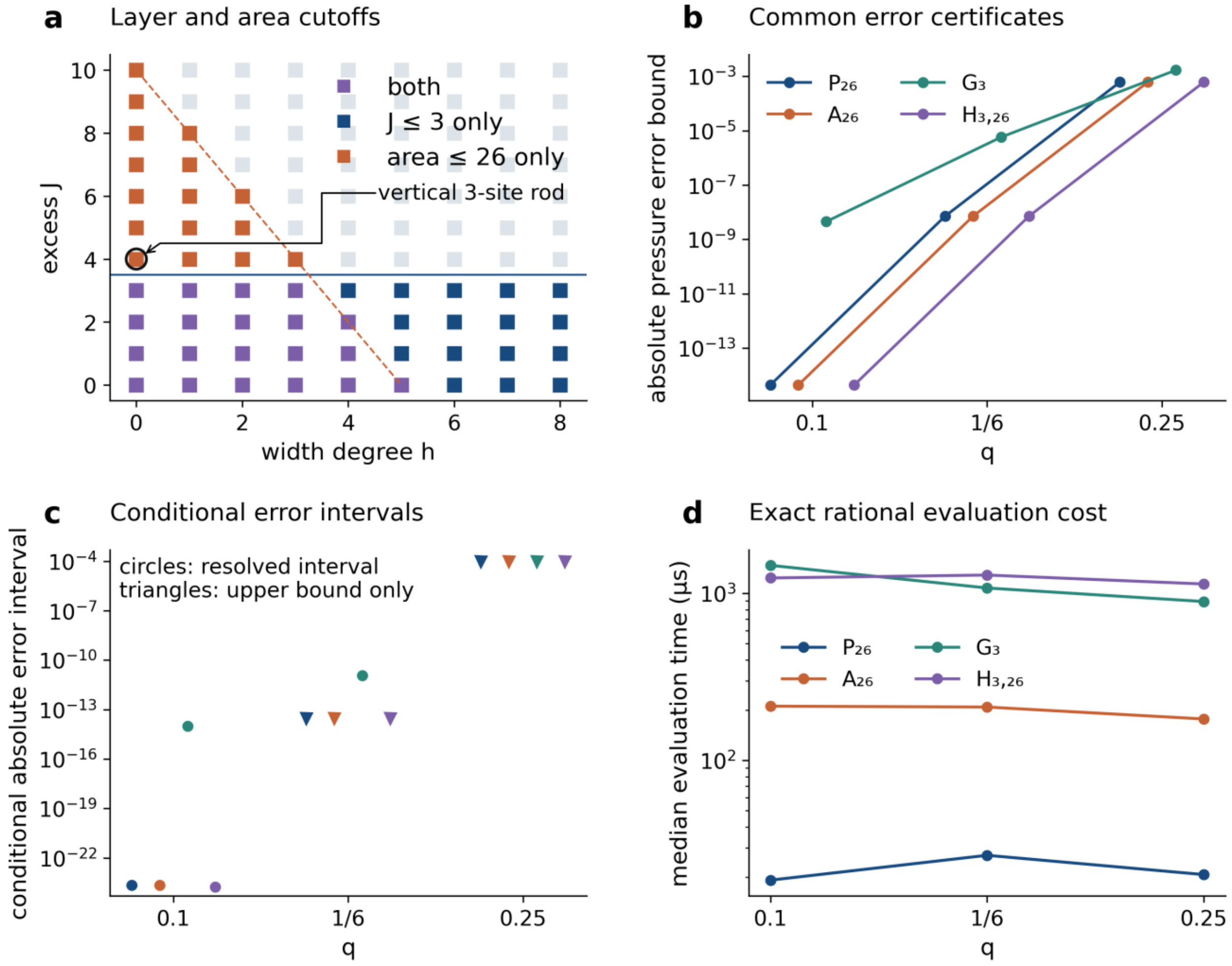

Figure 3. Pressure approximation depends on the retained set and the error evidence. (a) Lattice of width degree and total excess; color distinguishes the $J \leq 3$ and area $\leq 26$ cutoffs. The circled point is the first missing vertical rod. (b) Common absolute-error certificates, including repeat tails. (c) Conditional absolute-error intervals derived from the published series through $q^{52}$ and the contour norm. Circles denote intervals with a positive lower endpoint; triangles give upper bounds only. Colors match panel b. (d) Median warmed exact-rational evaluation times in the included measurement environment. Timings compare evaluation of already available formulas and coefficients, not their discovery or complete generation cost.